\documentclass[twocolumn,prl,showpacs,preprintnumbers,amsmath,amssymb]{revtex4-1}
\usepackage{graphicx}

\begin{document}

\title{Self-Energy Effects on the Low- to High-Energy Electronic Structure of SrVO$_3$}

\author{S. Aizaki$^1$, T. Yoshida$^1$, K. Yoshimatsu$^2$,
M. Takizawa$^{1*}$, M. Minohara$^2$, S. Ideta$^1$, A.
Fujimori$^1$, K. Gupta$^3$, P. Mahadevan$^3$, K. Horiba$^{2,5}$,
H. Kumigashira$^{4,5}$, M. Oshima$^{2,5}$}

\affiliation{$^1$Department of Physics, The University of Tokyo,
Tokyo 113-0033, Japan}

\affiliation{$^2$Deptartment of Applied Chemistry, The University
of Tokyo, Tokyo 113-0033, Japan}

\affiliation{$^3$S. N. Bose National Centre for Basic Sciences,
Kolkata 700-098, India}

\affiliation{$^4$KEK, Photon Factory, Tsukuba, Ibaraki 305-0801,
Japan}

\affiliation{$^5$JST-CREST, Tokyo 102-0075, Japan}

\affiliation{$^*$Present address : Research Organization of
Science and Engineering, Ritsumeikan University, 1-1-1
Noji-Higashi, Kusatsu, Shiga 525-8577, Japan}
\date{\today}

\begin{abstract}
The correlated electronic structure of SrVO$_3$ has been
investigated by angle-resolved photoemission spectroscopy using
{\it in-situ} prepared thin films. Pronounced features of band
renormalization have been observed: a sharp kink $\sim 60$meV
below the Fermi level ($E_F$) and a broad so-called ``high-energy
kink" $\sim$0.3 eV below $E_F$ as in the high-$T_c$ cuprates
although SrVO$_3$ does not show magnetic fluctuations. We have
deduced the self-energy in a wide energy range by applying the
Kramers-Kronig relation to the observed spectra. The obtained
self-energy clearly shows a large energy scale of $\sim$ 0.7 eV
which is attributed to electron-electron interaction and gives
rise to the $\sim$0.3 eV ``kink" in the band dispersion as well as
the incoherent peak $\sim$1.5eV below $E_F$. The present analysis
enables us to obtain consistent picture both for the incoherent
spectra and the band renormalization.
\end{abstract}

\maketitle
In a correlated electron system, coupling of single-particle
excitations with collective excitations leads to a pronounced
energy-dependent band renormalization, so-called kink, in the band
dispersion. In the studies of high-$T_c$ cuprate superconductors
by angle-resolved photoemission spectroscopy (ARPES), a ``kink"
has been observed around $\sim 60$ meV below $E_F$ \cite{Lanzara}
in the nodal region of the Fermi surface, where no superconducting
gap opens. Moreover, a ``high-energy kink" has also been observed
around $0.3$-$0.4$ eV below $E_F$ \cite{Graf,Meevasana} and its
origin has been debated. Electron-phonon interaction
\cite{Lanzara}, antiferromagnetic fluctuations, and/or the
magnetic resonance mode \cite{Johnson} have been proposed as
possible origins of the low energy kink. As for the high energy
kink \cite{Graf}, short-range Coulomb interaction
\cite{Meevasana}, a disintegration of an electron into a spinon
and a holon high-energy spin fluctuations \cite{Scalapino},
loop-current fluctuations \cite{Zhu} and electron correlation
\cite{Byczuk} have been proposed as a possible candidate. In order
to clarify the origin of the high energy kink, studies of kinks in
transition-metal oxides other than the cuprates will give useful
information.

One of the perovskite-type light transition-metal oxides (TMOs)
SrVO$_3$ (SVO) is a prototypical Mott-Hubbard-type system with the
$d^1$ electronic configuration and is an ideal systems to study
the fundamental physics of electron correlation. In fact, a
dynamical mean-field-theory (DMFT) calculation \cite{DMFT_RMP} of
the momentum-dependent spectral function of SVO has foreseen the
existence of a high-energy kink  caused by a general property of
electron-electron interaction and hence a general feature of the
electron self-energy of correlated metals \cite{Nekrasov}. To
address the nature of electron correlation in SVO, photoemission
spectroscopy measurements have been extensively performed
\cite{Imada, Fujimori, Inoue_PRB, Inoue_PRL, Sekiyama, Eguchi,
Maiti_Euro}. The V 3$d$ band dispersion and the Fermi surfaces of
bulk SVO were studied by ARPES by Yoshida {\it et al.}
\cite{Yoshida_PRL}. They obtained the mass enhancement factor of
$m^{*}/m_{b} \sim 2$ near $E_F$, consistent with the bulk
thermodynamic properties \cite{Inoue_PRB}. A more recent ARPES
study of SrVO$_3$ and CaVO$_3$ showed that the bandwidth indeed
decreased by $\sim 20$ \% in going from SrVO$_3$ to CaVO$_3$
\cite{Yoshida_PRB}. Takizawa {\it et al.} \cite{Takizawa}
fabricated SVO thin films having atomically flat surfaces using
the pulsed laser deposition (PLD) technique and studied its
detailed electronic structure by {\it in-situ} ARPES measurements.
Clear band dispersions were observed not only in the coherent
quasi-particle (QP) part but also in the incoherent part,
consistent with the DMFT calculation \cite{Nekrasov}.

In the present work, we have investigated the existence or absence
of the low and high energy kinks in SVO with improved sample
quality and instrumental resolution. A pronounced effect of
energy-dependent band renormalization, namely a ``kink" has been
observed at the binding energy of $\sim 60$ meV and a ``high
energy kink" at $\sim$0.3 eV. Since SVO is a Pauli-paramagnetic
metal without any signature of magnetic fluctuations, the presence
of the kinks will give us a clue to understand the nature of the
interaction which gives rise to the kinks. Furthermore, we deduced
the self-energy $\Sigma(\mathbf{k},\omega)$ in a wide energy range
through Kramers-Kronig analysis.

\begin{figure}
\includegraphics[width=8.6cm]{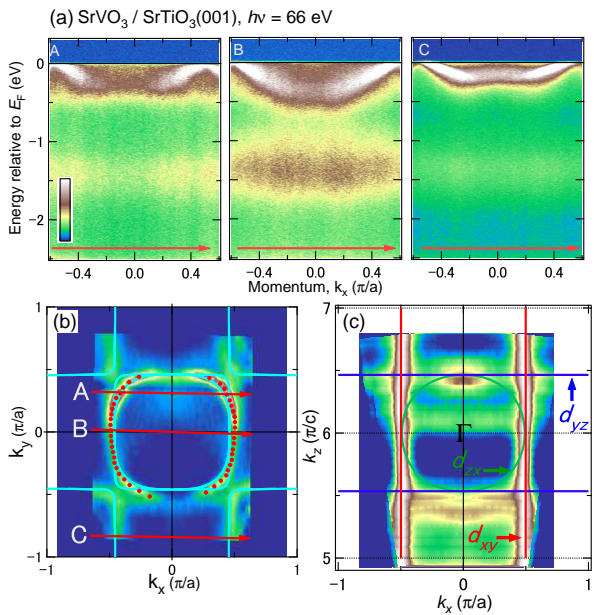}
\caption{\label{EDC_FS}(Color online) ARPES spectra of SrVO$_3$.
(a) Intensity plots in $E$-$k$ space for cuts A, B and C shown in
panel (b) taken at the photon energy of $66$ eV with linear
polarization. (b) Intensity map at $E_F$ revealing the nearly
circular cross-section of the $d_{xy}$ Fermi surface. Dots show
the $k_F$ points determined by MDC peak positions at $E_F$. Solid
curves are the FSs given by the LDA band-structure calculation.
(c) Spectral weight mapping in $k_{x}$-$k_{z}$ space obtained
using photon energies $h\nu$= 50-106 eV. Tight-binding bands
fitted to the Fermi surfaces are shown by circular and straight
lines. }
\end{figure}

{\it In-situ} photoemission measurements in the valence-band and
core-level regions were performed at beamlines 28A and 2C of
Photon Factory (PF) using a Scienta SES-2002 electron analyzer at
photon energies $h\nu$ = $50$ to $106$ eV and $h\nu$=800 eV,
respectively. Epitaxial thin films of SVO were grown on
single-crystal Nb-doped SrTiO$_3$ (001) substrates by the PLD
method. The substrates were annealed at 1050$^{\circ}$C under an
oxygen pressure of $\sim 1\times10^{-6}$ Torr to obtain an
atomically flat TiO$_2$-terminated surface. SVO thin films were
deposited on the substrates at 900$^{\circ}$C under a high vacuum
of $\sim 10^{-8}$ Torr. The chemical composition was checked by
core-level photoemission and the surface morphology by {\it
ex-situ} atomic force microscopy, showing atomically flat
step-and-terrace structures. ARPES measurements were performed in
an ultrahigh vacuum better than $1\times10^{-10}$ Torr below 20 K.
In-plane electron momenta $k_x$ and $k_y$ are expressed in units
of $\pi/a$, where $a = 3.905$ \textrm{\AA} is the in-plane lattice
constant of the SVO thin film, identical to that of the SrTiO$_3$
substrate. Electron momentum in the out-of-plane $k_z$ direction
is expressed in units of $\pi$/c, where $c = 3.82$ \textrm{\AA} is
the out-of-plane lattice constant of the SVO thin film determined
by x-ray diffraction reciprocal space mapping.

In SVO, each of the $d_{xy}$, $d_{yz}$ and $d_{zx}$ orbitals forms
a nearly two-dimensional band. Consequently, the Fermi surfaces
(FSs) consist of three cylinders penetrating perpendicularly to
each other. Figure \ref{EDC_FS}(a) shows ARPES spectra near $E_F$
along the cuts in the Fermi surface mapping of Fig.
\ref{EDC_FS}(b). There are two main features near $E_F$; the
coherent part (the sharp QP peak within $\sim 0.5$ eV of $E_F$)
and the broad incoherent part (often regarded as the remnant of
the lower Hubbard band centered $\sim 1.5$ eV below $E_F$). As
reported before \cite{Yoshida_PRL, Takizawa}, the coherent part
shows a clear band dispersion.

In order to examine the three-dimensional electronic structure, we
obtained the $E_F$ intensity map in $k_{x}$-$k_{z}$ space by
changing the photon energy as shown in Fig. \ref{EDC_FS}(c). Here,
the $k_z$ values have been obtained by assuming the inner
potential of $V_0$ = 18 eV and the work function of $\phi$ = 4.5
eV. The intensity distribution indicates that the $d_{xy}$ FS is
nearly a straight cylinder along the $k_z$ direction and is
therefore two dimensional. On the other hand, the $d_{zx}$ and
$d_{yz}$ FSs are not clearly observed due to matrix-element effect
and $k_z$ broadening.

\begin{figure}
\includegraphics[width=8.6cm]{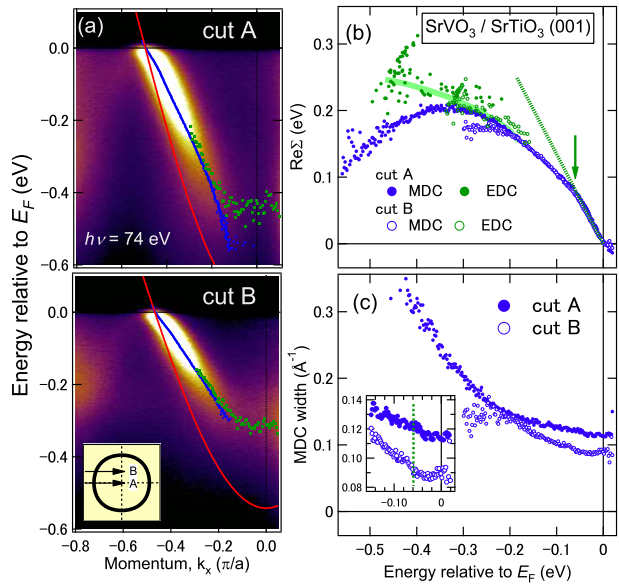}
\caption{\label{kink}(Color online) Band dispersions and
self-energies in the vicinity of the Fermi level. (a) Intensity
plots for cuts shown in inset. QP band dispersions are determined
by the MDC peak positions (blue dots), which are correct near
$E_F$, and the second derivative of EDCs (green dots), which are
correct near the band bottom. The non-interacting band given by
the LDA band-structure calculation is shown by red curves. (b)
Real part of the self-energy Re$\Sigma(\omega)$. Here, we have
obtained Re$\Sigma(\omega)$ as the difference between the band
determined by the experimental band dispersion and the LDA energy
band dispersion. The position of the kink at $\sim 60$ meV is
shown by an arrow. (c) Energy dependence of the MDC width, which
is proportional to imaginary part of the self-energy
Im$\Sigma(\omega)$. Inset is enlarged plots near $E_F$, which also
indicates the kink at $\sim 60$ meV.}
\end{figure}

 We shall investigate correlation effects in the QP
spectra by close examination of the nearly two-dimensional
$d_{xy}$ band. Image plots of the $d_{xy}$ band in $E$-$k_{x}$
space are shown in Fig. \ref{kink}(a). Here, QP band dispersions
are determined by the peak positions of the momentum distribution
curves (MDCs) and the second derivative of the energy distribution
curves (EDCs). Note that the spectral intensity at the bottom of
the dispersion is suppressed due to matrix-element effect. Hence,
the dispersion near the band bottom is well represented by EDC
peaks, while the MDC peaks well represent that near $E_F$. By
smoothly connecting the MDC peak dispersion and the EDC dispersion
in an intermediate energy region of $\sim$0.25 eV, one can obtain
a reasonable picture of the $d_{xy}$ band dispersion in the entire
energy range.

In order to deduce the real part of the self-energy
$\Sigma(\mathbf{k},\omega)$, we use the non-interacting band
dispersion calculated within the local-density approximation (LDA)
(red curve) and take the difference between the band dispersions
and the LDA band as shown in Fig. \ref{kink}(b). The nearly
identical $\Sigma(\mathbf{k},\omega)$ for cuts A and B indicates
that the self-energy $\sum$ is nearly $\mathbf{k}$-independent at
least within the studied momenta. In the deduced ${\rm
Re}\Sigma(\mathbf{k},\omega)$, a ``kink" is seen around $60$ meV
below $E_F$ and shall be referred to as the ``low-energy kink",
very similar to those observed in the high-$T_c$ cuprate
superconductors. As shown in Fig. \ref{kink}(c), the signature of
the kink is also seen in the MDC width, which is proportional to
${\rm Im}\Sigma(\omega)$. Lanzara {\it et al.} related the kink in
the high-$T_c$ cuprates with the oxygen half-breathing phonon mode
of $\sim 60$ meV \cite{Lanzara}. In SrTiO$_3$ (STO), which has the
perovskite-type crystal structure like SVO, an optical phonon mode
has been identified at $\sim 60$ meV \cite{Jiaguang}. Also, SVO
does not have low energy spin fluctuations unlike the cuprates.
Therefore, the present observation of the low-energy kink in SVO
can be unambiguously attributed to coupling of electrons to the
oxygen breathing modes.

The ${\rm Re}\Sigma(\omega)$ thus deduced shows not only the
low-energy kink ($\sim -60$ meV) but also a weak, broad ``kink" at
-0.3--0.4 eV, again similar to the high-energy kink of the
high-$T_c$ cuprates. Here, it should be noted that if MDC peaks
are used to deduce ${\rm Re}\Sigma(\omega)$ in the entire energy
range, the high-energy kink feature is over-emphasized compared to
the actual high-energy kink. Therefore, the correct deduction of
the experimental band dispersion using both MDC and EDC peaks is
necessary to study the high-energy kink phenomena.

\begin{figure*}
\includegraphics[width=17.8cm]{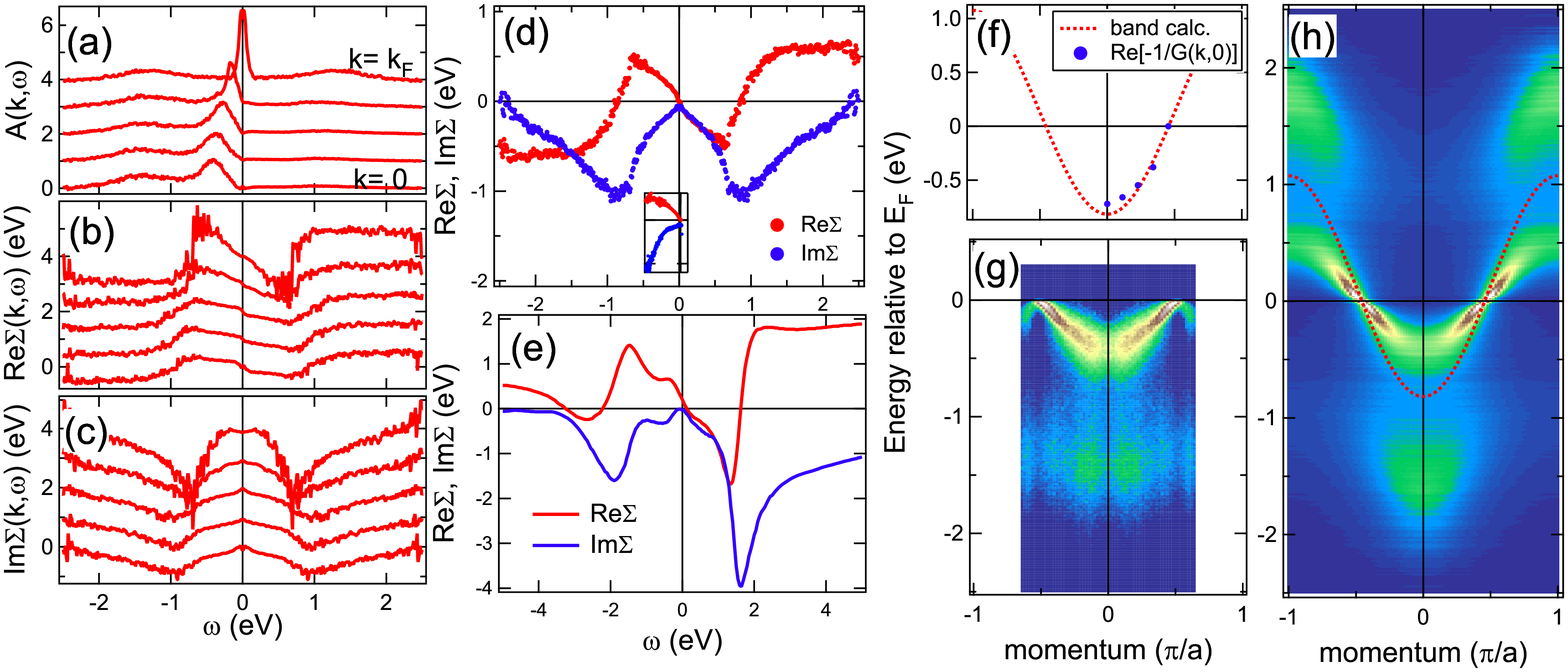}
\caption{\label{selfenergy}(Color online) Self-energy iteratively
deduced from the measured ARPES spectra using the Kramers-Kronig
transformation. (a): $A(k,\omega)$ calculated using the
self-energy in panels (b) and (c). (b) and (c): Re$\Sigma$ and
Im$\Sigma$ plotted as a function of energy for various momenta.
(d) Average of Re$\Sigma$ and Im$\Sigma$ over the momentum. For
comparison, the Re$\Sigma$ and Im$\Sigma$ (shifted by 0.3 eV)
obtained in Fig. \ref{kink} are shown in inset. (e) Self-energy
predicted by the LDA+DMFT calculation of Ref. \cite{NekrasovDMFT}.
(f) Comparison of $\mathrm{Re}[-1/G(k,0)]$(=$\epsilon_k$) obtained
from the experimental data with the LDA band dispersion
\cite{Pavarini}. (g) Intensity plot of the ARPES data. (h)
Simulation of the spectral function
$A(k,\omega)=-\mathrm{Im}G(k,\omega)/\pi$ using the
momentum-averaged self-energy in panel (d) and the LDA energy
band.}
\end{figure*}

The energy range of ${\rm Re}\Sigma(\mathbf{k},\omega)$ and ${\rm
Im}\Sigma(\mathbf{k},\omega)$ studied by the above method is
limited to $\sim$0.5 eV below $E_F$, the energy range of the
coherent part, while the behavior of the self-energy over a wider
energy range is necessary to understand the role of electron
correlation or the incoherent part, too. Therefore, we deduce the
self-energy in a wider energy range using the Kramers-Kronig (KK)
relation as follows. By performing the KK transformation of the
spectral function
$A(\mathbf{k},\omega)=-\mathrm{Im}G(\mathbf{k},\omega)/\pi$, one
can obtain $\mathrm{Re}G(\mathbf{k},\omega)$ and consequently,
$G(\mathbf{k},\omega)=1/(\omega-\epsilon_k-\Sigma(\mathbf{k},\omega))$
and hence $\epsilon_k+\Sigma(\mathbf{k},\omega)$. However, the
ARPES intensity only gives $A(\mathbf{k},\omega)$ bellow $E_F$.
Therefore, in the present analysis, several assumptions have been
made to deduce the self-energy. First, we assume electron-hole
symmetry for the self-energy, which can be justified near $E_F$:
$\mathrm{Re}\Sigma(\mathbf{k},\omega)=
-\mathrm{Re}\Sigma(\mathbf{k},-\omega)$ and
$\mathrm{Im}\Sigma(\mathbf{k},\omega)=
\mathrm{Im}\Sigma(\mathbf{k},-\omega)$. To obtain the self-energy
 self-consistently under this assumption, we use the experimental ARPES intensity
 $I(\mathbf{k},\omega)$ and construct the initial function for $A(\mathbf{k},\omega)$ as
$A(\mathbf{k},\omega)=I(\mathbf{k},\omega) (k<k_F)$,
$A(\mathbf{k},\omega)=I(\mathbf{k},\omega)+ I(\mathbf{k},-\omega)
(k=k_F)$. Here, $A(\mathbf{k},\omega)$ ($k<k_F$) for $\omega > 0$
is assumed to be much smaller than those for $\omega < 0$ because
the band dispersion $\epsilon_k$ is below the $E_F$. By
KK-transforming $A(\mathbf{k},\omega)$,
$\mathrm{Re}\Sigma(\mathbf{k},\omega)$ and
$\mathrm{Im}\Sigma(\mathbf{k},\omega)$ are obtained. In order to
fulfill the electron-hole symmetry, the
$\Sigma(\mathbf{k},\omega)$ for $\omega >0$ is set equal to the
complex conjugate of -$\Sigma(\mathbf{k},\omega)$ for $\omega<0$.
Then, $A(\mathbf{k},\omega)$ can be renewed by using the new
$\Sigma(\mathbf{k},\omega)$. This process is repeated iteratively
until $A(\mathbf{k},\omega)$ and $\Sigma(\mathbf{k},\omega)$ are
converged. Note that the converged $A(\mathbf{k},\omega)$ is
nearly identical to $I(\mathbf{k},\omega)$ for $\omega<0$,
indicating that the rather drastic assumptions were reasonably
realistic and the resulting self-energy is a good approximation
for the true self-energy in the wide energy range of a few eV.

The input $A(\mathbf{k},\omega)$ is taken from cut B in Fig.
\ref{EDC_FS} and the integrated background and the tail of the O
2$p$ band have been subtracted. Thus obtained
$A(\mathbf{k},\omega)$ for $k\leq k_F$ is shown in Fig.
\ref{selfenergy}(a). Panels (b) and (c) show $\mathrm{Re}\Sigma$
and $\mathrm{Im}\Sigma$, respectively, derived from
$A(\mathbf{k},\omega)$ in panel (a). The deduced
$\Sigma(\mathbf{k},\omega)$s for each momentum show line shapes,
but there is a weak momentum dependence. Particularly, the
absolute value of the $\Sigma(\mathbf{k},\omega)$ for $k=k_F$ is
larger than those for other momenta. Assuming that the momentum
dependence of $\Sigma$ is symmetric with respect to $k_F$, the
average of the self-energy over the Brillouin zone has been
obtained as shown in Fig. \ref{selfenergy}(d). The low energy part
of the iteratively deduced self-energy qualitatively agrees with
the self-energy deduced from the band dispersion as shown in the
inset of Fig. \ref{selfenergy}(d).

The non-interacting band dispersion $\epsilon_k$ is related to
$G(\mathbf{k},\omega)$ through
$\mathrm{Re}[-1/G(\mathbf{k},0)]=\epsilon_k$. Figure
\ref{selfenergy}(f) demonstrates good agreement between the
experimentally deduced $\mathrm{Re}[-1/G(\mathbf{k},0)]$ and the
band dispersion predicted by the band-structure calculation
\cite{Pavarini}. Interestingly, this analysis demonstrates that
the non-interacting band can be extracted from experimental data
which are influenced by electron correlation. One can therefore
conclude that it is self-consistent to use the band-structure
calculation for $\epsilon_k$ in the present analysis. In Figs.
\ref{selfenergy}(g) and \ref{selfenergy}(h), we compare the ARPES
intensity plot and the simulated intensity distribution
$A(\mathbf{k},\omega)$ in $E$-$k$ space. Here, we have calculated
$A(\mathbf{k},\omega)$ using the momentum-averaged
$\Sigma(\omega)$ [Fig.\ref{selfenergy}(d)] and the result of the
band-structure calculation $\epsilon_k$ \cite{Pavarini}. The QP
dispersions and the incoherent part shown in Fig.
\ref{selfenergy}(g) are successfully reproduced by the simulation
as shown in Fig. \ref{selfenergy}(h), indicating the
self-consistency of the deduced self-energy.

The $\mathrm{Re}\Sigma$ and $\mathrm{Im}\Sigma$ iteratively
deduced from the experiment exhibit the characteristic behaviors
predicted by the DMFT study of the Hubbard model \cite{Byczuk}.
For example, $\mathrm{Re}\Sigma$ shows a maximum at $\omega\sim$
-0.7 eV where the EDC shows an intensity minimum. In Fig.
\ref{selfenergy} (d) and (e), we compare the experimental
self-energy and the self-energy calculated for SVO by the LDA +
DMFT method \cite{NekrasovDMFT}. The experimental self-energy
shows remarkably similar behavior to that of the LDA + DMFT,
although both the vertical and horizontal axes are different by a
factor of two [see panels (d) and (e)].

As for the ``high energy kink", while the theory predicts a
pronounced feature $\sim$ 0.3 eV below $E_F$, the same energy
region as the present experiment although the experimental kink is
less pronounced. The kink which we observed at $\sim$60 meV (Fig.
\ref{kink}) is not reproduced by the calculation because
electron-phonon coupling is not included in the theoretical model.
As for the incoherent part, while the peak energy of the LDA +
DMFT calculation is $\sim$2.1 eV below $E_F$ \cite{NekrasovDMFT},
that in the present result is $\sim$1.5 eV below it, suggesting
that electron correlation is overestimated in the calculation. On
the other hand, both the QP band in experiment and that from the
calculation show a renormalization factor of $\sim$2 relative to
the LDA band dispersion. This is because the slope of ${\rm
Re}\Sigma(\omega)$ near $E_F$ which is nearly the same between the
experimental and the theoretical self-energies as shown in Figs.
\ref{selfenergy}(d) and (e). Therefore, in order to achieve a
consistent picture of the correlated electronic structure, further
development of the theoretical approaches and the extension of the
experimental approach to other correlated systems have to be made
in future studies.

In conclusion, we have performed a detailed ARPES study of {\it
in-situ} prepared SVO thin films and revealed self-energy effects
on the QPs as well as the incoherent structure. A low energy kink
was observed at the binding energy $\sim$ 60 meV as in the case of
the high-$T_c$ cuprate superconductors, and is attributed to
electron-phonon coupling. We have obtained the self-energy in a
wide energy range by applying the KK relation to the experimental
spectral function. The self-energy shows a large energy scale of
$\sim$ 0.7 eV reflecting electron-electron interaction and giving
rise to the broad ``high-energy kink" $\sim$0.3 eV below $E_F$ as
well as the incoherent peak $\sim$1.5 eV below $E_F$. The present
result provides a self-consistent procedure to experimentally
deduce the self-energy in correlated electron systems and this
procedure would be useful for future studies of electron
correlation effects.

 The authors would like to thank A. Georges and S. Biermann for
discussion and K. Ono, J. Adachi and M. Kubota for their support
in the experiment at KEK-PF. This work was supported by a
Grant-in-Aid for Scientific Research (S)(22224005), a Grant-in-Aid
for Young Scientist (B) (22740221), a Grant-in-Aid for Scientific
Research on Innovative Area "Materials Design through Computics:
Complex Correlation and Non-Equilibrium Dynamics" and an
Indo-Japan Joint Research Project ``Nobel Magnetic Oxide
Nano-Materials Investigated by Spectroscopy and ab-initio
Theories" from JSPS. The work in Kolkata was supported by DST,
India. The experiments were done under the approval of Photon
Factory Program Advisory Committee (Proposals No. 2007G597 and No.
2008S2-003) at the Institute of Material Structure Science, KEK.

\bibliography{SVOthinfilm}

\end{document}